\documentstyle[mprocl]{article}

\def\be{\begin{equation}}
\def\ee{\end{equation}}
\def\bea{\begin{eqnarray}}
\def\eea{\end{eqnarray}}
\begin{document}

\title{ From Reissner-Nordstr\"om quantum states \\ to 
charged black holes mass evaporation}

\author{
 P.V. Moniz}

\address{DAMTP, 
University of Cambridge\\ Silver Street, Cambridge, CB3 9EW, UK
\\{\small e-mail: {\sf 
prlvm10@amtp.cam.ac.uk; 
paul.vargasmoniz@ukonline.co.uk}}
\\{\small WWW-site: {\sf 
http://www.damtp.cam.ac.uk/user/prlvm10}}}

\maketitle\abstracts{In this report we describe 
 quantum  
Reissner-Nordstr\"om (RN) black-holes 
interacting with a complex scalar field. 
Our analysis  is characterized  
by solving a Wheeler-DeWitt 
equation in the proximity of an apparent horizon of the RN space-time. 
Subsequently, we obtain 
a wave-function $\Psi_{{\rm RN}}[M, Q]$ representing 
the RN black-hole. A special 
emphasis is given to the evolution of 
the mass-charge rate  affected  by  Hawking radiation.
More details can be found in ref. \cite{mpla}.}

Recently, there has been a renewed interest 
in the canonical quantization of black-hole 
space-times \cite{pol}$^{\!-\,}$\cite{mpla}. 
The general 
aim is to obtain a  description 
of quantum black holes that would 
 go beyond a semi-classical 
approximation. 
In this report, we will extend M. Pollock's 
method \cite{pol} (which was itself influenced by  
the   work of A. Tomimatsu \cite{tomi}) to 
 Reissner-Nordstr\"om (RN) black-hole. 
As a consequence, we will  find a  wave function for the RN 
 black-hole, 
which will have an explicit dependence on 
its  mass $M$ and the charge $Q$. 
The T.O.H. (Tomimatsu-Oda-Hosoya)
method\cite{odat,odasolo} 
constitutes another very interesting and similar 
approach to this purpose.

The main elements 
of the Hamiltonian formalism used here  follow  ref. \cite{thomhaji}. 
{\rm  En route} towards our reduced model  we further 
take the following steps. To begin with, 
our 4-dimesional {\em spherically symmetric}
 metric is written as 
\begin{equation}
ds^2 =  h_{ab} dx^a dx^b + \phi^2 ( d\theta^2 + \sin^2\theta d\varphi^2 ),
\label{eq:jap2}
\end{equation}
together with the ADM decomposition 
\begin{equation}
h_{ab} = \left(
\begin{array}{cc}
-\alpha^2 + \frac{\beta^2}{\gamma} & \beta \\
\beta & \gamma
\end{array}\right) ~,~ 
h^{ab} = \left( 
\begin{array}{cc}
-\frac{1}{\alpha^2} & \frac{\beta}{\alpha^2\gamma} \\
\frac{\beta}{\alpha^2 \gamma} & 
\frac{1}{\alpha} - \frac{\beta^2}{\alpha^2\gamma^2}
\end{array}
\right),
\label{eq:jap3}
\end{equation}
where $\alpha, \beta, \gamma, \phi$ are functions 
of $(x^0, x^1) = (\tau, r)$ which will be defined later 
(see eq. (\ref{eq:vad2})). 
We also take a scalar field 
$\hat \psi = \psi (\tau, r)$ and the 
electromagnetic field 
$
F_{ab}  =  \varepsilon_{ab} \sqrt{ -h} E; ~E = ( - h)^{-1/2} 
(\dot A_1 - A_0^\prime), $ $
\hat D_a \hat \psi  =   \partial_a \psi + ieA_a\psi; 
\hat D_2 \hat \psi = \hat D_3 \hat \psi = 0. $
Notice   we employ $``\cdot'' \equiv \frac{\partial}{\partial \tau}$ 
and $``\prime'' \equiv \frac{\partial}{\partial r}$.
The next {\em significant}  step consists in  choosing  the 
following 
coordinate gauge: $
\alpha = \frac{1}{\sqrt{\gamma}} \Leftrightarrow \sqrt {- h} =1$.

The constraints 
of our model will be expressed 
  in terms of dynamical quantities defined at  an 
apparent horizon of the RN black-hole, 
which is defined by the condition\cite{thomhaji}: $
h^{ab}(\partial_a \phi)(\partial_b \phi) = 0 \Leftrightarrow 
\dot{\phi}(\dot{\phi} - \phi^\prime) = 0.$
In order to obtain a satisfactory 
description of an evaporating RN black 
hole on the apparent horizon, it is more 
convenient to use a RN - Vaidya metric \cite{pol}.
With
 $v \equiv  \tau + r = t + r^*$ as the 
advanced null Eddington-Finkelstein coordinate, and 
$r^*$ as the corresponding ``tortoise" 
coordinate,   the relationship between 
the time $\tau$ and the time coordinate 
$t$ for the standard  RN metric is 
$\tau = t - r + r^*$. The 
Vaidya RN metric becames
\begin{equation}
ds^2 = -\left(1 - \frac{2M}{r} 
+ \frac{Q^2}{r^2}\right)d\tau ^2 
+ \left(\frac{4M}{r} 
- \frac{2Q^2}{r^2}\right)d\tau dr 
+ \left(1 + \frac{2M}{r} - 
\frac{Q^2}{r^2}\right)d r^2 + 
\phi^2d\Omega _2^2.
\label{eq:vad2}
\end{equation}
The RN black-hole has two apparent horizons, namely at 
$
r_{\pm}(v)  = M (v) \pm $ $ \sqrt{M^2(v) - Q^2}$ 
and we will henceforth restrict ourselves to 
the case of $r_+$. 
In similarity with the Schwarzschild case 
we also  take $M \simeq M(\tau)$ in the vicinity of the apparent horizon. 
After some lenghty calculations
we obtain 
(see ref. \cite{mpla} for more details)
 the approximate expressions: 
\begin{eqnarray}
{\cal H}_0 & = & \frac{1}{2}\pi_\phi - 
\frac{1}{8} + \frac{Q^2}{32M^2} - 
\frac{1}{16\pi }\frac{1}{\rho^2}(\pi ^2_{\chi } 
+ \pi_{\psi_1}^2) - \rho^2\pi (\psi_1^{\prime^2} +  
 \chi^{\prime 2}) \nonumber \\
& - &  \frac{1}{2\rho^2}\pi _{A_1}^2 - 
\frac{\rho^2}{2}[2\pi e^2 A_1^2 (\psi_1^2 
+ \chi^2) ] - 2 \pi e \rho^2 A_1 (\psi_1\chi^\prime - 
\chi \psi_1^\prime )]
\label{eq:H0b}\\
{\cal H}_1 & = &\frac{1}{2}\pi_\phi - 
\frac{1}{8} - \frac{3Q^2}{32M^2} + 
\frac{1}{2}[\pi_{\psi_1} \psi_1^\prime + 
\pi_{\chi} \chi^\prime] + \frac{e A_1}{4}
(\pi_{\psi_1} \chi  - \pi_\chi \psi_1),  
\label{eq:H1b}
\end{eqnarray}
with $
\psi  =  \sqrt{2\pi }(\psi_1 + i\chi ), 
\pi_{\psi }  \rightarrow  \frac{1}{\sqrt{8\pi }} \pi_\psi$.
Compatibility\cite{mpla}
requires that 
$
\pi_{\psi _1} = -4\pi \rho^2\psi_1^\prime, ~ 
\pi_\chi = - 4 \pi \rho^2 \chi^\prime,$
with the terms with $ A_1$ to  be 
negligible, and with the conditions $A_0 = 0$ together   with 
$\dot \psi = \dot \psi^{\dagger} = 0$. 
We further take $A_1 = A_1 (r)$.

Quantization proceeds via the operator 
replacements
\begin{equation}
\pi_{\phi } \rightarrow  - 
i\frac{\partial }{\partial \phi } \simeq  
-i\frac{2M^2}{4M^2 + Q^2}\frac{\partial }{\partial M},
\pi_{\psi_1} \rightarrow  - 
i\frac{\partial }{\partial \psi _1},
\pi_{\chi } \rightarrow  - 
i\frac{\partial}{\partial \chi },
\end{equation}
which yelds the Wheeler-DeWitt equation 
for the wave function $\Psi$,
\begin{equation}
- i 
\frac{2M^2}{4M^2 + Q^2} \frac{\partial \Psi}{\partial M}= - 
\frac{1}{16\pi M^2 + \pi Q^4/M^2 - 8\pi Q^2} 
\left[\frac{\partial^2 \Psi }{\partial \psi_1^2} 
+ \frac{\partial^2 \Psi }{\partial \chi^2} \right ] 
+ \frac{1}{4}\Psi,
\label{eq:wdw3}
\end{equation}
whose solutions are 
\begin{equation}
\Psi_{{\rm RN}} [
M, Q; \psi_1, \chi; k]  = \Psi^0_{{\rm RN}} 
e^{i\left [\frac{1}{4}\left (- \frac{Q}{M} 
+ 2M\right ) + k^2\frac{M}{Q^2 - M^2}
\pm 2\sqrt {\pi }k(\psi _1 + \chi ) \right]}
\zeta (A_1),
\label{eq:solt1}
\end{equation}
where $k^2$ is a separation constant and 
$\Psi^0_{{\rm RN}}$ an integration constant.

It is interesting to notice the following as well. For the Schwarzschild case 
($Q=0)$ eq. (\ref{eq:solt1}) implies that near to $M=0$ the wave 
function will oscilate with infinite frequency. If $\dot{M} < 0$,  this would 
represent the quantum mechanical behaviour of the black hole 
near the end point of its evaporation. In the RN case, the 
rapid oscillations 
will occur again for $M=0$ but also when 
$M \sim Q$. I.e., near extremality and when the black hole 
mass evaporation 
can eventually  stop. Hence, the presence of $Q$ in $\Psi_{{\rm RN}}$ 
allowed us to identify some known   physical situations of the 
RN bkack hole.

As far as the mass-charge ratio  for the RN 
black hole is concerned, we
get the equation
\begin{equation}
\dot{M} + 
\frac{1}{4M^2} \frac{a[k^2; Q]}{d(M)} + 
\frac{1}{4M^4} \frac{b[k^2; Q]}{d(M)}
+ \frac{c[k^2; Q]}{d(M) M^6}
= 0,
\label{eq:newlast1}
\end{equation}
where
\begin{equation}
a[k^2; Q]  =  k^2 - 5Q^2 + Q; 
b[k^2; Q]  =  k^2 Q^2 - 3Q^4; 
c[k^2; Q]  =  \frac{k^2 Q^4}{8}; 
d(M)  =  1 + \frac{Q^2}{2M^2}.
\end{equation}
An  integration of (\ref{eq:newlast1}) leads to the
result
\begin{equation}
M = \left[ M_0^3 - \frac{3}{2}(k^2 - 5 Q^2 + Q)\right]^{1/3} 
(t - t_0)^{1/3} .
\label{eq:mrate2}
\end{equation}
We can now identify several physical cases of interest for the RN 
black hole, according if $a, b, c, d$ are either positive, zero or negative.

For the case of $Q=0$ \cite{pol} (Schwarzschild),  it is the separation 
constant $k^2$ that determines if the black hole is evaporating and 
decreasing its mass ($k^2 >0$), or increasing its mass 
($k^2 < 0 \Leftrightarrow k$ imaginary). In the present RN 
black hole case, the presence of the charge $Q$ introduces significative 
changes.
A $\dot{M} > 0$ stage can be obtained, 
with $k^2 >0$ but $a<0, b<0$ and $c>0, d>0$. If $Q=0$, this possibility is 
absent. When $Q\neq 0$, $k^2 < 0$, then $c<0, b<0, a<0$, if 
$1 + 20 k^2 > 0$ has real solutions.

Overall, our results do bring additional 
information regarding quantum black-hole,  but there is still 
the need for further investigation.

\section*{Acknowledgments}
This work was supported by the JNICT/PRAXIS XXI Fellowship 
BPD/6095/95. The author is  grateful to M. Cav\`aglia, 
A. Hosoya,  J. Louko and M. Ryan   for interesting conversations and 
 discussions, as well as to 
 J. M\"akel\"a  for further important comments and suggestions.
 He also wishes to thank I. Oda 
for letting him know of his recent work and for extensive correspondence.

\end{document}